\documentclass[]{spie}  %>>> use for US letter paper
\usepackage[]{graphicx}

\title{On-sky tests of sky-subtraction methods for fiber-fed spectrographs} 

\author{Myriam Rodrigues \supit{a}, M. Cirasuolo \supit{c,d}, F. Hammer \supit{b}, F. Royer \supit{b}, C. J. Evans \supit{d}, M. Puech \supit{b}, H. Flores \supit{b},  I. Guinouard \supit{b}, Gianluca Li Causi \supit{d}, K. Disseau \supit{b}, Y. Yang \supit{b}. 
\skiplinehalf
\supit{a} European Southern Observatory, Alonso de Cordova 3107, Santiago, Chile; \\
\supit{b} GEPI , Observatoire de Paris, CNRS, University Paris Diderot \\ 5 Place Jules Janssen, 92195 Meudon, France;\\
\supit{c} Institute for Astronomy of University of Edinburgh\\  Royal Observatory, Blackford Hill, EH9 3HJ, Edinburgh, U.K.;\\
\supit{d} UK Astronomy Technology Center, Blackford Hill, EH9 3HJ, Edinburgh, U.K.;\\
\supit{e} INAF, Rome Astronomical Observatory, Viale del Parco Mellini 84, 00136 Rome Italy;\\
}

\authorinfo{Further author information: Send correspondence to Myriam Rodrigues,  marodrig@eso.org}
\begin{document} 
  \maketitle 

%%%%%%%%%%%%%%%%%%%%%%%%%%%%%%%%%%%%%%%%%%%%%%%%%%%%%%%%%%%%% 
\begin{abstract}

We present preliminary results on on-sky test of sky subtraction methods for fiber-fed spectrograph. Using dedicated observation with FLAMES/VLT in I-band, we have tested the accuracy of the sky subtraction for 4 sky subtraction methods: mean sky, closest sky, dual stare and cross-beam switching. The cross beam-switching and dual stare method reach accuracy and precision of the sky subtraction under 1\%. In contrast  to the commonly held view in the literature, this result points out that fiber-fed spectrographs are adapted for the observations of faint targets. 
\end{abstract}

%>>>> Include a list of keywords after the abstract 


%%%%%%%%%%%%%%%%%%%%%%%%%%%%%%%%%%%%%%%%%%%%%%%%%%%%%%%%%%%%%

%%-----------------------------------------------------------
\section{Context}
\label{sec:title}
%%-----------------------------------------------------------
%%  Use following command to specify that graphics file is in 
%%  a directory other than this LaTeX source file.
%%  Note use of / to separate subdirectories, for UNIX and Windows OS.
%%\graphicspath{{H:/HANSON/SPIESTY/}}
%% tabular environment useful for creating an array of images  
Historically,  fiber-fed spectrographs had been deemed inadequate for the observation of faint targets, because of two main reasons: (1) the low throughput of fibers that implies a low global efficiency of the spectrograph; (2) the difficulty to achieve high accuracy on the sky subtraction. However, thanks to the important progress in fiber technology, modern fibers spectrographs now have global efficiencies close to multi-slit instruments. To date, the main drawback had been the quality of the sky subtraction, critical for the observation of faint targets. The impossibility to sample the sky in the immediate vicinity of the target in fiber instruments has led to a commonly held view that a multi-object fibre spectrograph cannot achieve an accurate sky subtraction under 1\% contrary to their slit counterpart. In the past years, severals studies have focused on the subject and several designs of sky subtraction strategies to properly subtract the sky have been proposed [\citenum{1992MNRAS.257....1W},\citenum{2005NewA...11...81L},\citenum{2008MNRAS.386...47E},\citenum{2010MNRAS.408.2495S},\citenum{2010SPIE.7735E.216R}]: 
\begin{itemize}
\item \textbf{Simultaneous sky fibers} - Several fibers are dedicated to sampling the sky in the region of observations. The number of sky fibers depends on the wavelength domain observed, the dimension of the field of view and the required quality of sky-subtraction. Observations have to be previously corrected from the individual response of the fiber and scattered light. The mean sky among all sky fibers or the closest sky fiber is subtracted to the fiber object spectrum.
\item \textbf{$\lambda-\lambda$ reconstruction}This method is a variant of the previous one [\citenum{2010SPIE.7735E.216R}]. Dedicated sky fibers are distributed in the field of view. The sky continuum is reconstructed at different wavelengths in all the FOV by interpolation and the sky emission lines are subtracted using the algorithm of Davies, 2009 [\citenum{2007MNRAS.375.1099D}].
\item \textbf{Dual fiber bottom in stare mode}. The sky is sampled simultaneously in the immediate vicinity of the object by a sky fiber associated to the object fiber. This setup is close to a pseudo slit. This configuration implies to dedicate half of the fibers to the sky measurement. 
\item  \textbf{Beam Switching} - Each fiber is switched between an object and a sky position. This strategy has the advantage to observe the sky background with the same fibers used for the targets [e.g. \citenum{1992MNRAS.257....1W}]. The sky and the instrumental signal can be subtracted simultaneously. However, this procedure implies to dedicate half of the observation time to sky sampling.
\item \textbf{Cross Beam Switching} - Each science target has two fibers associated, as in the dual fiber mode. The object is observed in the fibers following a sequence A B B A or A B A.  Despite the drawback of decreased spectrograph multiplex capability, the dual fiber button design has the advantage to be similar to chopping into a slit and thus is 100\% of the time on the scientific targets and allows a very accurate sky and instrumental response subtraction. 
\end{itemize}
 
We show hereafter the capabilities of four of these methods - mean sky, closest sky, dual stare and cross-beam switching - in the I-band using as a benchmark dedicated FLAMES/GIRAFFE/VLT observations. A similar assessment is ongoing in the near-IR regime using FMOS observations.

%-------------

\section{Observation data}

We have undergone a technical observation in FLAMES/GIRAFFE/VLT in order to test the capabilities of the different methods for subtracting the sky in fiber-fed instruments. FLAMES/GIRAFFE is a multi-object, intermediate and high resolution spectrograph at VLT. We use the MEDUSA mode that allows to observe 132 separated object in a 25' diameter field of view.  Each fiber has an aperture of 1.2 arcsec on the sky.

\subsection{Description of the technical observations}
 
Observations have been taken on the March 8th 2012 with FLAMES/GIRAFFE in Medusa Mode. The fibers have been distributed in a 20'x20' region of the zCosmos field as follow: 31 as singles fibers - 26 with obj and 5 with sky - and 70 fibers in pairs. Each pair has two fibers separated by 12" and oriented N-S, a \textit{South} fiber and a \textit{North} fiber,  see Figure 1. Three pairs are 'pure sky fibers pairs' without object in any of the fibers and 31 have the object located in the southern fiber. The observations have been carry out using the cross beam-switching configuration: the telescope has been offset by 12" in the north direction in 3 consecutive A-B sequences. As FLAMES does not have a beam switching mode, a ABAB sequence has been chosen, instead of the commonly used ABBA, in order to minimize the risk of loosing the objects in the fibers during the two B exposures when secondary guiding is not available. The error in pointing due to the offset is 0.2", which is smaller than the diameter of the fibers. \\
The exposure time was set to 600s for each individual exposure and the reddest setting LR09 (central lambda 8817.00\AA\, and R=6500). An attached-flat has been taken just after the end of the sequence.  The field has been observed when crossing the meridian (HA $\pm$ 30min) and at low airmass ($<1.2$). The background sky from the moon was extremely strong, a full moon was at $\sim$28 degrees from the field. This corresponds to 50\% of the background light arising from the moonlight. 
 
 %-------------
   \begin{figure}
   \begin{center}
   \begin{tabular}{c}
   \includegraphics[height=9cm]{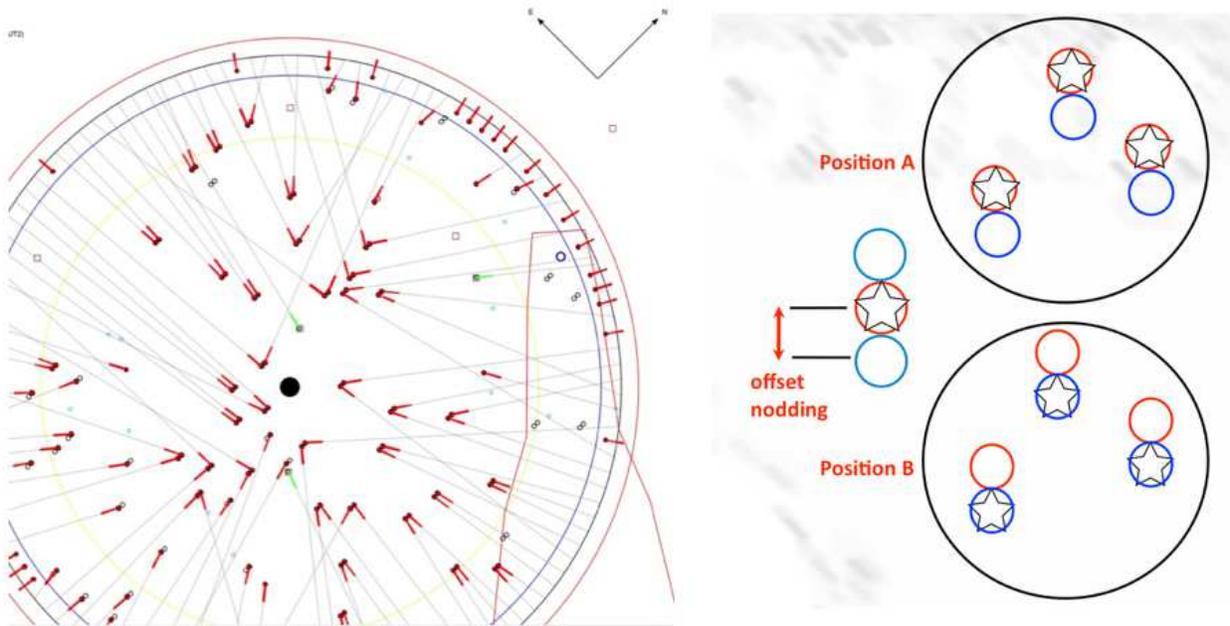}
   \end{tabular}
   \end{center}
   \caption[example] 
%>>>> use \label inside caption to get Fig. number with \ref{}
   { \label{fig:example} 
Figure 1 - \textbf{Left}: Plate configuration of the GIRAFFE observations.  \textbf{Right}:  Dual fibers configuration and the cross-switching mode. Notice that the pair \#9 has systematically higher residual for all the methods. This pair may be affected by a faint object or by contamination by neighbor fibers. }
   \end{figure}

 \subsection{Data reduction}
 
 The observations have been set to the configuration of the cross beam-switching mode. The other modes have been "simulated" by using different  data reduction strategies. The standard data reduction has been done using the ESO pipeline ( bias, lamp-flat, wavelength calibration and extraction of the spectra).  
 
  %\begin{figure}
   %\begin{center}
   %\begin{tabular}{c}
   %\includegraphics[height=9cm]{A-B_frames.eps}
   %\end{tabular}
   %\end{center}
   %\caption[example] 
%>>>> use \label inside caption to get Fig. number with \ref{}
   %{ \label{fig:example} 
%Left: Raw A frame. Right: A-B frame. Objects in emission and continuum are clearly visible. The sky lines are directly subtracted. The dark traces correspond to fibers with object in B position - negative of the object. }
   %\end{figure}
   
 \begin{itemize}
\item Simultaneous sky fibers - Frames A and B have been combined separately and the spectra have been extracted in the combined-A  and -B frames. The sky has been subtracted by using the median sky on all the available sky fibers ( dedicated sky fibers and fibers in sky position from the fiber pairs). The sky subtracted spectra in A and B position are then combined. The sky-subtraction using the closest sky fiber has also been tested. In this case the closest sky fiber, not in the pair fiber, has been used.
%\item $\lambda-\lambda$ reconstruction - Such as the previous method, the A and B frames have been combined separately and spectra extracted. $N_\#$ sky fibers have been randomly picked. The surface of the sky continuum in the fov has been calculated at different wavelength and the interpolated sky continuum spectrum in the position of the target has been subtracted to object spectrum. The sky emission lines have been then subtracted using the method of Davies 2009. The detail description of the algorithm can be found in Rodrigues et al. 2010. 
\item Dual fiber bottom in stare mode. Spectra have been extracted in combined-A and B frames. The sky subtraction has been done by subtracting the fiber in Sky position to the associated Object fiber in the pair. The sky-subtracted spectra in A and B position are then combined.  The final spectrum for one target is composed by the combination of the spectra in fiber~$South$ in $A$ with those of the fiber~$North$ in $B$. 
\item  Cross Beam Switching - B frames have been subtracted to the previous A frames. The (A-B) frames have been combined in a final frame and the spectra have been extracted. The extracted spectra in (A-B) and (B-A) frames are finally combined.  The standard beam switching mode corresponds to the same extraction but without combining the (A-B) and (B-A) spectra (half of the exposure time on the object compared to the other methods). 
\end{itemize}

 \subsection{Sky subtraction accuracy estimators}

The accuracy of the sky subtraction has been evaluated using the 3 pure-sky fiber pairs and defined as following: 
 \begin{equation}
Res(\lambda)= \frac{Obj(\lambda) - Sky(\lambda)}{<Sky(\lambda)>}
\end{equation}
where $<Sky(\lambda)>$ is the median sky in all the sky-fibers. The signal $Res(\lambda)$ is composed by two components: a low frequency signal which depends on the quality of the subtraction of the sky background and of the instrumental response; and a high frequency signal induced by the read-out noise (\textit{ron}). As our aim is to evaluate the accuracy of the sky subtraction, the high frequency signal has been filtered using a wavelet decomposition algorithm [\citenum{1994A&A...288..342S}]. $Res(\lambda)$ has been decomposed into 6 wavelet scales: the 5th first scales correspond to the high frequency signal of the \textit{ron} and the last scale corresponds to the low frequency scale. Only the regions free of strong sky lines have been taken into account to estimate the smoothed residual. The quality of the sky subtraction has been evaluated by calculating the average value (accuracy) and the rms (precision) of the last wavelet scale of the smoothed $Residual(\lambda)$ signal. 

  \section{Sky subtraction accuracy}
 
Figures 2 shows the residuals for the three pure-sky-fiber pair using the cross beam-switching method. This method gives the best results, with accuracy and precision of the sky subtraction under 1\%. The dual stare mode gives similar results. The closest and mean sky methods give accuracy and precision in the sky subtraction ten times lower, see Figure 3. The accuracy and precision of the sky subtraction in each of the pure sky fibers are tabulated as a function of  the used sky subtraction method in table 1. 

 \begin{figure}
   \begin{center}
   \begin{tabular}{c}
   \includegraphics[height=9cm]{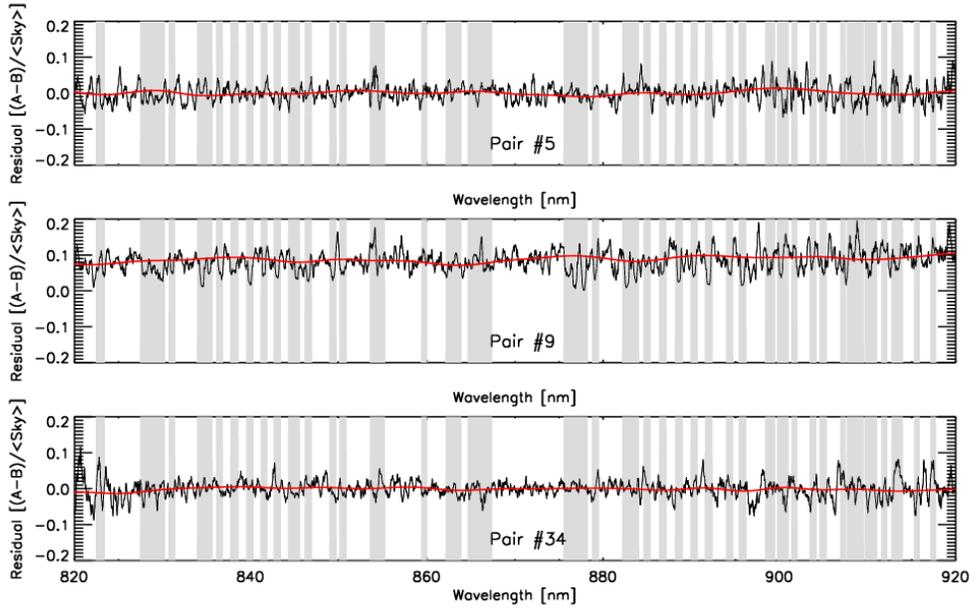}
   \end{tabular}
   \end{center}
   \caption[example] 
%>>>> use \label inside caption to get Fig. number with \ref{}
   { \label{fig:example} Residual sky plotted in solid black line for one of the three pure sky fiber pairs using the nodding dual method. The regions of the spectrum affected by strong sky emission lines are delimited in light grey area and have not been taken into account when measuring the accuracy and precision of the sky subtraction. The smoothed residual, without read-out noise, is plotted in red.  }
   \end{figure}

 \begin{figure}
   \begin{center}
   \begin{tabular}{c}
   \includegraphics[height=9cm]{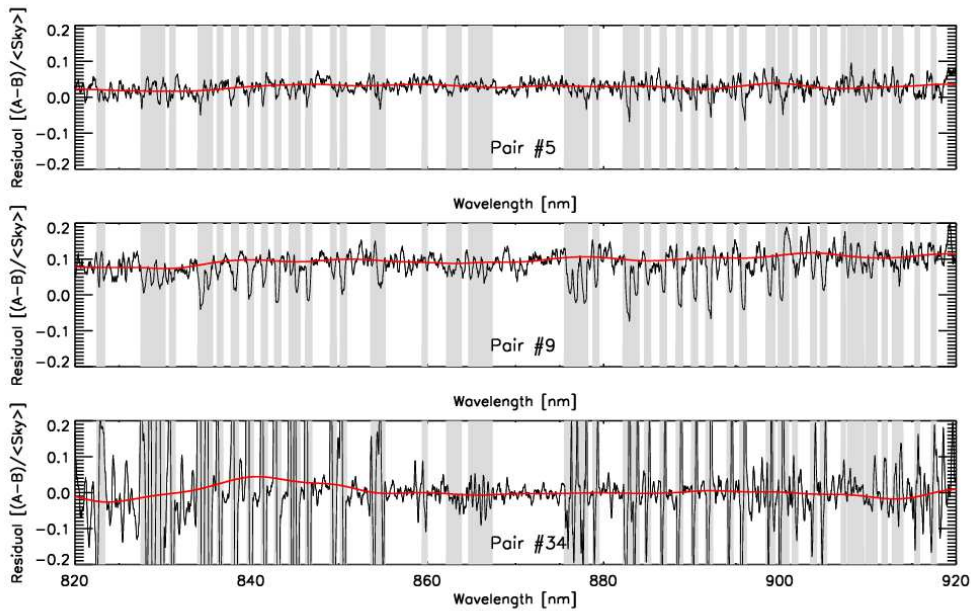}
   \end{tabular}
   \end{center}
   \caption[example] 
%>>>> use \label inside caption to get Fig. number with \ref{}
   { \label{fig:example} The same than Figure 2 but with the mean sky subtraction method. }
   \end{figure}
   
\begin{table}[htdp]
\caption{Accuracy and precision of the sky subtraction for four methods in the three pure sky pairs}
\begin{center}
\begin{tabular}{|c|c|c|c|}
\hline
$N_\# pair$& Method&Accuracy& Precision\\
\hline
&Mean sky&0.0695&0.0945\\
Pair 5&Closest sky&0.0295&0.0312\\
&Dual Stare&-0.0028&0.0077\\
&Cross BS&-0.0012&0.0059\\
\hline
&Mean sky&0.1600&0.2400\\
Pair 9&Closest sky&0.1011&0.1024\\
&Dual Stare&0.0969&0.0975\\
&Cross BS&0.0938&0.0949\\

\hline
&Mean sky&0.0374&0.0576\\
Pair 34&Closest sky&-0.0153&0.0373\\
& Dual Stare&-0.0002&0.0060\\
&Cross BS&-0.0025&0.0079\\
\hline
\end{tabular}
\end{center}
\end{table}%

The small difference between the quality of the sky subtraction using cross beam-switching and the dual stare method is surprising. The cross beam-switching method allows to correct precisely the instrumental response - such as scatter light, pixel-to-pixel variations and fiber transmission - but is affected by the temporal variation of the sky background between two nodding exposures. On the contrary, the dual stare method is not dependent on the temporal variation of the sky but is affected by the systematics from the instrumental response. Our study shows that these two methods reach similar quality of the sky subtraction in the spectral window $820-920nm$. This suggests that the instrumental systematics are correctly corrected, $<1\%$, with standard data reduction methods and that the temporal variation of the sky background in exposure $\sim600s$ is also under $<1\%$. The instrumental systematics seems less critical than what is commonly assumed in the literature [\citenum{1998ASPC..152...50W}].   

These conclusions are only valid for instruments with the same stability and characteristics that FLAMES-GIRAFFE (fraction of scatter-light, flexures, CCD characteristic ). In the case of the near-IR, the dual stare-mode is expected to be a better method because the instrumental systematics can be much larger ( due to the use of near-IR detectors). In addition, the temporal variation of the sky and the thermal background are stronger in the near-IR and it is thus required to have a faster beam-switching sequence. We are currently analyzing SUBARU/FMOS observations to test the several sky subtraction methods in the near-IR. \\

We illustrated, in Figure 4, the capabilities of the beam-switching and dual stare methods compared to the commonly used sky subtraction methods to detect  faint lines from distant galaxies. Lines as faint as $3.67\times10^{-17}\,erg/s/cm^2/Hz$ are detected after 1 hour of exposure in the dual stare mode. Note that this result is very promising since it corresponds to the worst case scenario: observation of a faint galaxy at few degrees $>30^o$ from the full moon. The gain in S/N on the detection of faint lines when using dual stare or beam-switching is clearly visible in the plots. The loss in multiplex when using these two methods is balanced by the high quality of the sky subtraction that allows to detect extremely faint objects. 
\begin{figure}
   \begin{center}
   \begin{tabular}{c}
      \includegraphics[height=10.5cm]{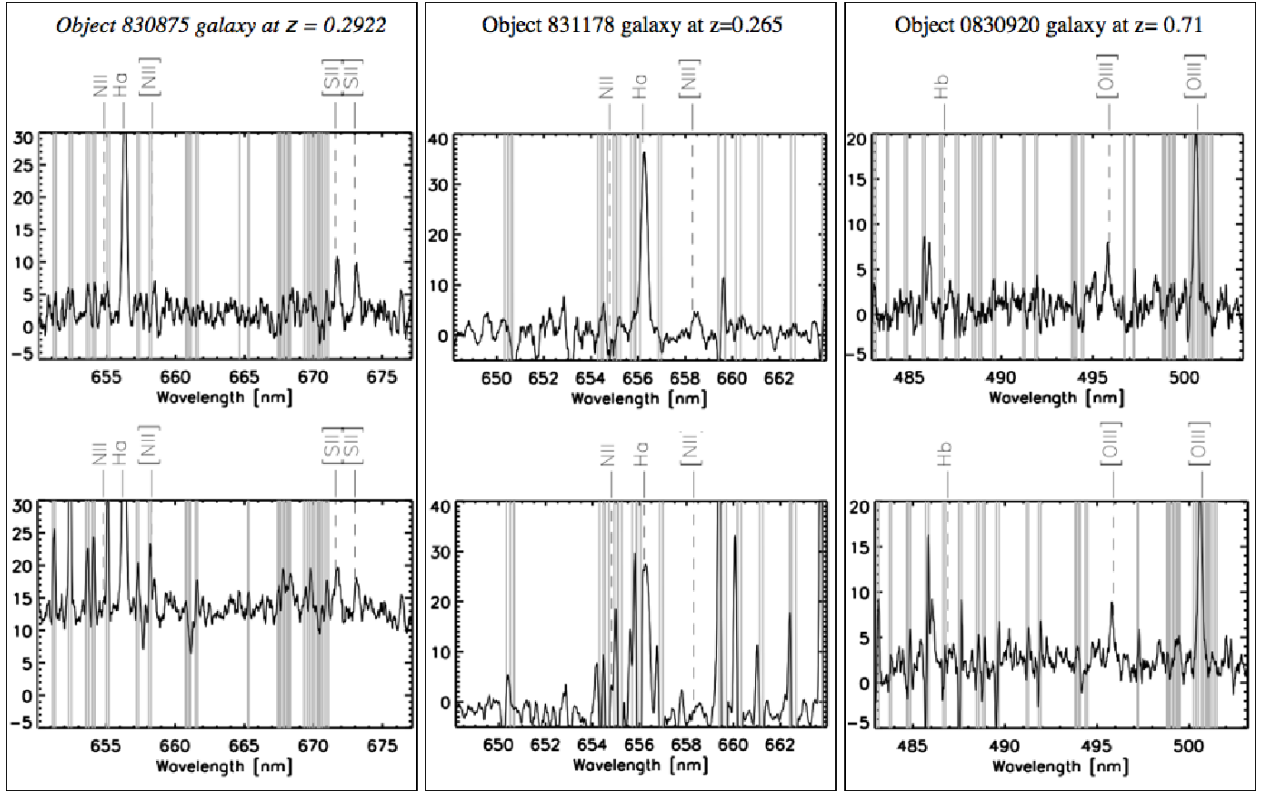}
   \end{tabular}
   \end{center}
   \caption[example] 
%>>>> use \label inside caption to get Fig. number with \ref{}
   { \label{fig:example} Comparison of the sky subtracted spectrum of faint lines galaxies using dual beam-switching (up) and the mean sky method (down). The dual stare method gives similar results to the cross beam-switching. The total exposure time is 3600s. Regions with sky lines are indicated in grey area. From left to right:   
 \textbf{Object 830875}:  galaxy at $z=0.2922$. $H\alpha$ at 848nm nm with a flux $4.31\times10^{-16} erg/s/cm^2/Hz$. The [SII] doubled falls at the 868 and 869 nm and have a flux equal to $5\times10^{-17}$ and $4.20\times10^{-17}erg/s/cm^2/Hz$ respectively. \textbf{Object 831178}: galaxy at 0.265. $H\alpha$ at 830.1nm with a flux $2.89\times10^{-16} erg/s/cm^2/Hz$. \textbf{Object 0830920}: galaxy at z$\sim0.71$: Three lines are visible $H\beta$ at 841nm, [OIII]4959 at 857.3 and [OIII]5007 at 840.4. Their respective flux are $3.67\times10^-{17}$, $5.21\times10^{-17}$ and $2.09\times10^{-16}erg/s/cm^2/Hz$.  }
   \end{figure}

 \section{Conclusion}
We present the preliminary results on on-sky test of different sky subtraction techniques for fiber-fed spectrograph. Using FLAMES/VLT observations, we have demonstrated that sky background can be subtracted within less that 1\% of residual, when using beam-switching or pairs fibers methods. This results point out that fiber-fed spectrographs are well suited for the observations of faint objects and that the issue of the sky subtraction is not a stopper for this technology.  A similar study is currently in progress to test the accuracy of the different sky subtraction method for near-IR spectrograph. Other sky subtraction methods, such as the $\lambda-\lambda$ reconstruction method [\citenum{2010SPIE.7735E.216R}] and a details study of the accuracy of the method as a function of the number of dedicated sky fibers are also under investigation. This study have been carry out in the framework of the design of MOONS/VLT [\citenum{2011Msngr.145...11C}] and OPTIMOS-EVE/E-ELT [\citenum{2011Msngr.145...11C},\citenum{2010SPIE.7735E..88N}] - two projects of multi-object spectrographs in the optical-to-nearIR domain based on fiber technology. 

\acknowledgments     %>>>> equivalent to \section*{ACKNOWLEDGMENTS}       
The authors are grateful to Andreas Kaufer,  Claudio Melo, Peter Hammersley and Suzanne Ramsay for approving and supporting our request for technical time in FLAMES/VLT. Thanks to Maja Vuckovic for carring out this observation at UT2/Paranal.

%%%%%%%%%%%%%%%%%%%%%%%%%%%%%%%%%%%%%%%%%%%%%%%%%%%%%%%%%%%%%
%%%%% References %%%%%

\bibliography{MOONS.bib}   %>>>> bibliography data in report.bib
\bibliographystyle{spiebib}   %>>>> makes bibtex use spiebib.bst

\end{document}